\documentclass[lettersize,journal]{IEEEtran}
\usepackage[T1]{fontenc}
\usepackage{amsmath,amsfonts}
\usepackage{algorithmic}
\usepackage{algorithm}
\usepackage{array}
\usepackage[caption=false,font=normalsize,labelfont=sf,textfont=sf]{subfig}
\usepackage{textcomp}
\usepackage{stfloats}
\usepackage{url}
\usepackage{verbatim}
\usepackage{graphicx}
\usepackage{cite}
\usepackage[utf8]{inputenc}
\usepackage[switch, modulo]{lineno}
\usepackage{hyperref}
\usepackage{xcolor}
\usepackage{tabularray}
\usepackage{mdframed}
\usepackage{makecell}
\usepackage{booktabs}
\usepackage{multirow}

\begin{document}

\title{TIARA: a fast gamma-ray detector for range monitoring in Proton Therapy}

\author{A. André, M. Pinson, C. Hoarau, Y. Boursier, M. Dupont, L. Gallin Martel, M.-L. Gallin Martel, A. Garnier, J. Hérault, J.-P. Hofverberg, P. Kavrigin, C. Morel, J.-F. Muraz, M. Pullia, S. Savazzi, D. Maneval and S. Marcatili

\thanks{This work did not involve human subjects or animals in its research.}
\thanks{A. André, C. Hoarau, L. Gallin Martel, M.-L. Gallin Martel, M. Pinson, P. Kavrigin, J.-F. Muraz and S. Marcatili are with Univ. Grenoble Alpes, CNRS, Grenoble INP, LPSC-IN2P3, 38000 Grenoble, France (e-mail: adelie.andre@lpsc.in2p3.fr, sara.marcatili@lpsc.in2p3.fr). }
\thanks{ Y. Boursier, M. Dupont, A. Garnier and C. Morel are with Aix-Marseille Univ, CNRS/IN2P3, CPPM, Marseille, France.}
\thanks{J. Hérault, J.-P. Hofverberg and D. Maneval are with Centre Antoine Lacassagne, 06200 Nice, France.}
\thanks{M. Pullia and S. Savazzi are with the CNAO Foundation, Milano, Italy}
}

%\markboth{Journal of \LaTeX\ Class Files,~Vol.~14, No.~8, August~2024}%

%\IEEEpubid{0000--0000/00\$00.00~\copyright~2024 IEEE}

\maketitle

\begin{abstract}
We developed a novel gamma-ray detection system (TIARA) for range monitoring in Particle Therapy. The system employs Cherenkov-based gamma-ray detection modules arranged around the target or patient, operated in time coincidence with a fast plastic beam monitor (described in a separate paper). This work focuses on the design and comprehensive characterization of the gamma-ray detection module.
It consists of a monolithic PbF$_2$ crystal (2~$\times$~1.5~$\times$~1.5~cm$^{3}$) coupled to a 2~$\times$~2~SiPM matrix from Hamamatsu. A series of beam tests at different clinical facilities (MEDICYC and ProteusOne in France, CNAO in Italy) enabled the determination of the detector time resolution under various conditions, with values ranging from 222 to 283~ps FWHM (Full Width Half Maximum). Monte Carlo simulations including the optical response of PbF$_2$ allowed for the determination of the detection efficiency as a function of particle type and energy. For the final TIARA prototype, which will be composed of 30~modules, an overall detection efficiency of 0.45\% is expected.
Comparison with experimental data confirmed that the modules are effectively insensitive to neutrons, yielding an excellent signal-to-noise ratio (SNR), with an estimated SNR of 17~for a module placed at 25~cm from the 148~MeV proton beam axis. These features translate into high range accuracy: while the performance varies with beam energy and irradiation conditions, a range accuracy of 3.3~mm at 2$\sigma$ significance level was achieved at low intensity with 63~MeV protons at MEDICYC, for a small irradiation spot of $\mathbf{\sim}$~10$\mathbf{^{7}}$~protons.
\end{abstract}


\section{Introduction}
\IEEEPARstart{P}{roton} therapy is an advanced cancer treatment modality that offers superior dose conformity to the tumor due to the characteristic energy deposition profile of charged particles, releasing the bulk of their energy at the end of their path (Bragg peak) \cite{Durante_2016}. While this physical advantage allows for better sparing of healthy tissues, it also introduces increased sensitivity to range uncertainties arising from patient mispositioning, inter-session anatomical variations, and inaccuracies in stopping power estimation derived from conventional X-ray CT imaging. 
Reducing the need for broad safety margins and improving treatment precision requires the development of reliable online proton range verification systems \cite{paush_2020, parodi_2018}.\\
Globally, several research efforts are underway to address this challenge \cite{kraan_2024}. As the primary proton beam is fully stopped inside the patient, monitoring techniques rely on the detection of secondary particles. Among these, Prompt Gamma (PG) emissions resulting from nucleon-nucleon collisions in the patient, have proven to be very promising due to their strong spatial and temporal correlation with the dose deposition profile \cite{min_2006, KRIMMER_2018, wronska_2020}. One such technique, Prompt Gamma Timing (PGT), measures the precise Time-of-Flight (TOF) between the proton entrance in the patient (a beam monitor provides the start signal) and the PG detection in a device placed downstream (stop signal)\cite{golnik_2014}. The feasibility of this approach has been demonstrated in several studies \cite{hueso_gonzalez_2015, werner_2019, pennazio_2022, Jacquet_2023,Heller_2024,ellin_2024,arino_2025}.\\
Building upon the PGT principle, we propose Prompt Gamma Time Imaging (PGTI) \cite{Jacquet_2021}, an innovative technique that goes beyond range verification. PGTI aims to reconstruct the full spatial emission profile of PGs by combining high-precision TOF measurements with a dedicated detector system called TIARA (Tof Imaging ARrAy). This capability opens the door to real-time, in vivo monitoring of anatomical variations, enabling the identification of discrepancies between the planned and delivered treatments. \\
%TIARA consists in a fast beam monitor \cite{andre_2024} read-out in time coincidence with a set of gamma-ray detection modules (8 in the current prototype, 30 in the final version) surrounding the anatomical region of interest. 
TIARA consists of a fast beam monitor \cite{andre_2024} operated in time coincidence with a set of 8 gamma-ray detection modules (current prototype)  arranged around the anatomical region of interest. In its clinical-scale version TIARA will include 30 gamma-ray modules. 
In order to reach a range accuracy of the order of a few mm (at 2$\sigma$), we have previously demonstrated with Monte Carlo simulation \cite{Jacquet_2021} and experiments \cite{Jacquet_2023} that a Coincidence Time Resolution (CTR, defined in section~\ref{sec:tres}) of the order of 100~ps RMS  (Root Mean Square) or 235~ps FWHM, is required. Moreover, since this technique is based exclusively on the measurement of particle TOF, the knowledge of the PG energy is not a requisite. \\
This paper focuses on presenting the design and implementation of the TIARA gamma-ray detection modules. Its full characterisation is achieved through a combination of experiments and Monte Carlo simulations. The module performances are reported in terms of: Detector Time Resolution (DTR) (section~\ref{sec:tres}), detection efficiency (section~\ref{sec:simu}), with a focus on background sensitivity (section~\ref{sec:back}), and range accuracy (section~\ref{sec:range}). 
 The results stem from two years of R\&D that we aim to present here in a structured manner. 
 %Minor differences between the tested prototypes can be attributed to the different stages of development reached at the time we tested them.
 %
\section{TIARA block detector and the case for Cherenkov radiators}\label{sec:det_des}
\begin{figure}[!h]
    \centering
    \includegraphics[width = 5 cm]{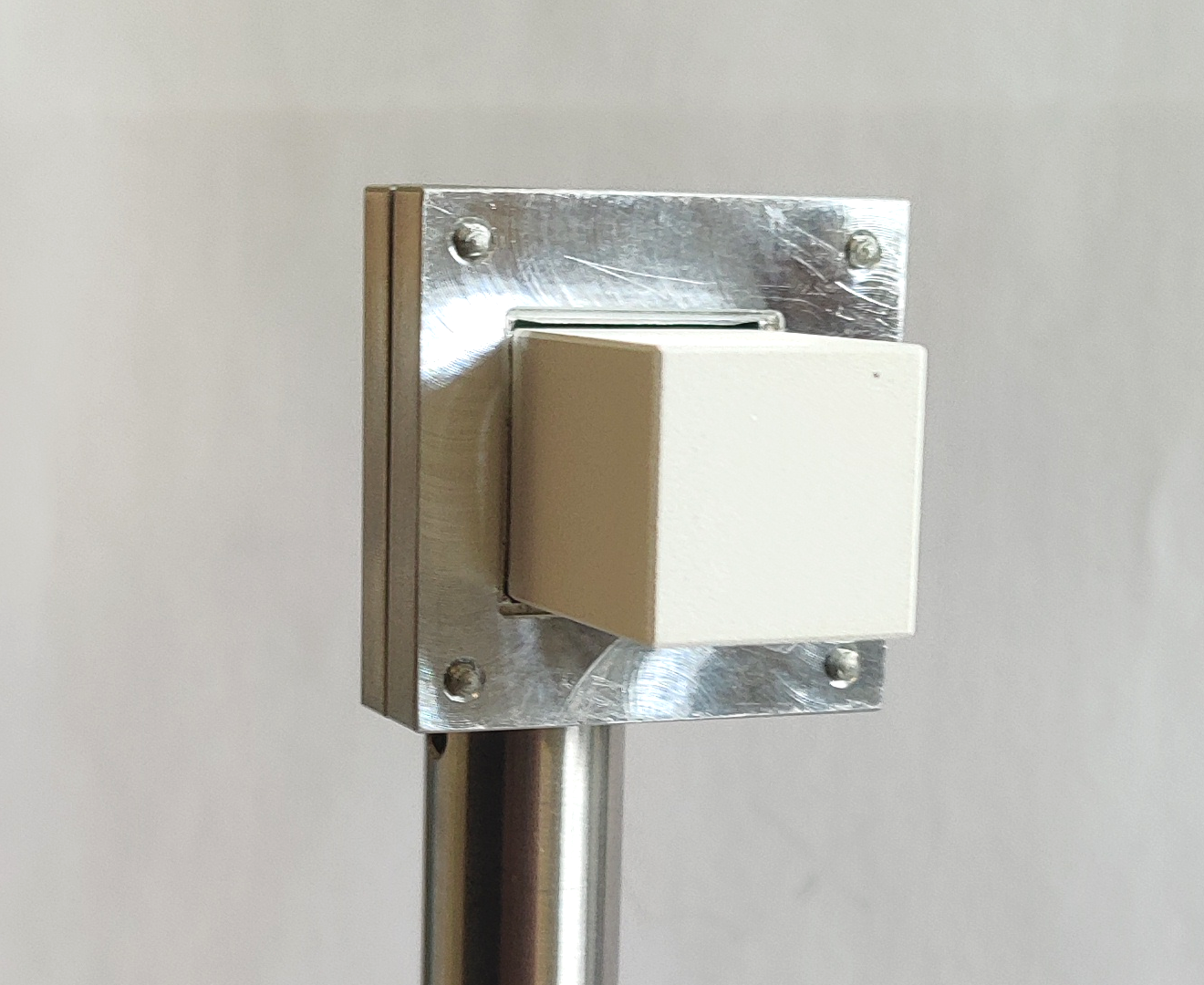}
    \caption{The TIARA block detector consists of a 2~$\times$~1.5~$\times$~1.5~cm$^{3}$ PbF$_2$ crystal covered by a white reflective paint and readout by a 2~$\times$~2~matrix of 6~$\times$~6~mm$^{2}$ SiPMs, multiplexed in a single readout channel.}
    \label{photo_detector}
\end{figure}

The ideal detector for PGTI should combine several key features: (i) $\approx$100~ps (rms) DTR, (ii) compact geometry to allow beam irradiations from different angles and ensure compatibility with intensity-modulated proton therapy (IMPT), (iii) high detection efficiency to compensate for the intrinsically low PG yield, and (iv) minimal sensitivity to background radiation (neutrons, electrons, and scattered protons), thereby optimizing the signal-to-noise ratio (SNR) and reducing pile-up under high-intensity conditions. \\
The time resolution of a gamma-ray detection module (typically composed of a converter and a photodetector) depends on three main factors: (i) the intrinsic timescale of the physical process generating optical photons in the converter, (ii) the temporal coherence of the emitted photons, and (iii) the number of photons reaching the photodetector, since higher statistics improves precision. The design of the TIARA detection modules seeks to optimize these three parameters simultaneously by relaxing the constraint of energy measurement.\\
The current TIARA detection module is based on a monolithic block detector design made of a 2~$\times$ 1.5~$\times$ 1.5~cm$^{3}$ PbF$_2$ crystal from Advatech. The rear face of the crystal is coupled to a 2~$\times$~2~SiPM matrix (6~$\times$~6~mm$^{2}$ S13360-6075PE pixels from Hamamatsu ) using the Dymax 3094 optical glue. The 4~SiPMs are readout with an hybrid layout front-end \cite{Cervi_2017, Nies_2018}, described in \cite{andre_2024}; it provides a parallel distribution of the bias voltage and a series connection of the output signal to optimize the time resolution by minimizing the overall capacitance. 
The other 5~crystal faces are covered with EJ-510 reflective paint to maximize the collection of Cherenkov photons. 
The complete detector is shown in Fig.~\ref{photo_detector} with the SiPMs and amplifier protected by an aluminium packaging. \\
With a limited crystal volume of few cubic centimeters, the optical photon propagation path is reduced and their temporal dispersion minimized. However, the full absorption of the incident PG seldom occurs, precluding reliable energy measurement.
At the same time, while even the fastest inorganic scintillators have characteristic scintillation decay times of several tens of nanoseconds, making them prone to pile-up at high counting rates, the Cherenkov process is intrinsically faster, with photon emission occurring at the picosecond scale. This enables unprecedented timing performance in gamma-ray detection, while also producing inherently short signals that are particularly advantageous under the high-rate conditions of clinical beams. Nevertheless, the choice of the converter material is crutial to fully exploit these potential advantages.\\
 According to the the Frank and Tamm equation~\cite{frank_1991}, the number of Cherenkov photons emitted per unit length and wavelength by a charged particle travelling faster than the speed of light in the medium ($\beta >1/n$) is given by:
 \begin{equation}
\frac{d^{2}N}{d\lambda dx}=\frac{2\pi \alpha z^{2}}{\lambda^{2}} \left(1-\frac{1}{\beta^{2} n^{2}(\lambda)}\right)
\label{cher_diff}
\end{equation}
with: $\lambda$, the Cherenkov photon wavelength; $\alpha$, the fine structure constant; $n$, the medium refractive index; $\beta$, the particle velocity relative to that of light; and $z$, the particle charge. For the detection of gamma-rays, Cherenkov emission is triggered by the secondary electron/positron either resulting from photoelectric absorption (mostly negligible for small volume detectors), Compton scattering or pair production, and therefore $z=1$, while charged particle energy is comprised between 2 and~10~MeV. If equation~\ref{cher_diff} is integrated over the working range of most common photodetectors (e.g., 350~nm~$< \lambda <$~550~nm), the number of Cherenkov photons emitted per unit range (of the charged particle) is obtained:
\begin{equation}\label{cher_int}
\frac{dN}{dx} \approx 476 \left(1-\frac{1}{\beta^{2} n^{2}}\right )~ \text{ph/cm}
\end{equation}
In high-Z media, electrons follow tortuous paths, leading to nearly isotropic photon emission rather than a well-defined Cherenkov cone. For MeV electrons, it results in only a few hundred optical photons, yielding a light output about two orders of magnitude lower than that of scintillators. This limited yield places strong constraints on timing optimization. Maximizing photon collection therefore requires a crystal with both a high refractive index and strong optical transparency, while maintaining a high Z to enhance PG detection probability. Among available materials, lead fluoride (PbF$_2$) is one of the best candidates, with $n \sim$ 1.8 (slightly varies with wavelength), high density (7.77~g/cm$^3$), and the highest transmittance among Cherenkov radiators~\cite{Mao_2012}.
Thallium chloride (TlCl) also offers similar properties~\cite{rebolo_2023}, while lead tungstate (PbWO$_4$) was excluded due to its lower transparency and residual scintillation component, which would complicate pure Cherenkov signal selection.\\
Most Cherenkov photons are emitted at short wavelengths ($\sim$300~nm or less) as $\frac{d^{2}N}{d\lambda dx}\sim \frac{1}{\lambda^{2}}$, requiring photodetectors with adequate UV sensitivity. While MicroChannel plate (MCP) detectors offer the best timing, their bulk makes them unsuitable for compact modules. For TIARA, Silicon Photomultipliers (SiPMs) provide the best compromise: compact, low-voltage, intrinsically fast (tens of ps), and with a Photo Detection Efficiency (PDE) larger than 30\% down to 350~nm in the case of S13360-6075PE devices.\\
Finally, it will be shown in section~\ref{sec:back} that pure Cherenkov radiators as PbF$_2$ do not allow the direct  detection of neutrons that are constituting the main source of background in particle therapy applications, thus offering large SNR.\\
%
%%%%%%%%%%%%%%%%%
\section{Time Resolution}\label{sec:tres}
\begin{figure}[!h]
    \centering
    \includegraphics[width = 8 cm]{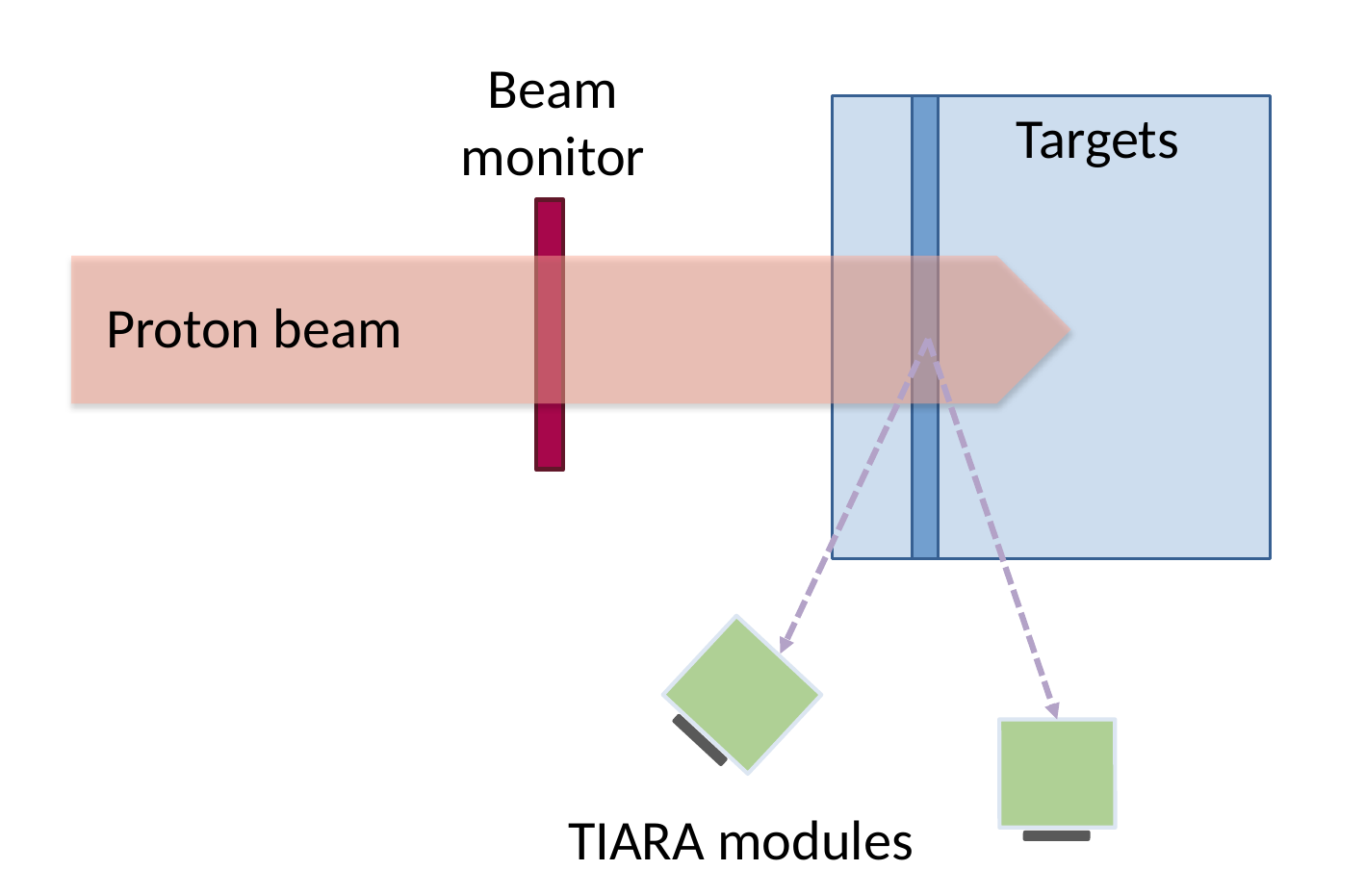}
    \caption{Experimental setup employed to measure the TOF  between the proton arrival time in the beam monitor and the PG detection in the TIARA module. For all experiments, the gamma module was placed upstream the target, while for the ProteusOne campaign a second module was positioned downstream at 90° with respect to the Bragg peak. A thin target (10~mm PMMA or 1~mm copper) is used for time resolution measurements, while a  thick (17~cm PMMA) target allows determining the SNR in more realistic conditions,  where the beam is fully absorbed. }
    \label{schema_type}
\end{figure}

%%%%%%%%%%%%%%%%%
%\subsection{Experimental conditions}
%
Time resolution is the key feature of TIARA as it directly impacts the system range accuracy in a treatment monitoring context. 
Its measurement is achieved through the irradiation with protons of a thin target (either a 10~cm thick PMMA slab or a 1~mm thick copper foil), which is acting as a point-like PG source. This hypothesis stands if the proton travel time in the target is well below the system time resolution.
In all experiments, the TIARA detector ($T_{stop}$) is placed downstream of the beam monitor, around the target (at different angular positions) and read in time coincidence with a beam monitor ($T_{start}$) based on a 25~$\times$~25~$\times$~1~mm$^{3}$ fast plastic scintillator readout by SiPMs \cite{andre_2024} that is placed upstream (cf. Fig.~\ref{schema_type}). The signals from both detectors are acquired by the WaveCatcher \cite{Breton_2014} digital sampler, with 500~MHz bandwidth and a 3.5~Gs/s sampling rate. Data analysis is performed offline and the time measurement is based on a digital Constant Fraction Discriminator (CFD) set at 10$\%$ of the maximum amplitude.
The system CTR corresponds to the standard deviation (or FWHM) of the measured TOF ($T_{stop}-T_{start}$) distribution and  results from the convolution of the gaussian Impulse Response Functions (IRFs) of the monitor and the gamma-ray detector. The associated variance (CTR$^2$) is equal to the sum of the IRFs' variances of the monitor  (DTR$^2_{m}$) and gamma-ray detector (DTR$^2_{\gamma}$):
CTR$^2$ = DTR$^2_{m}$ + DTR$^2_{\gamma}$.
The gamma-ray module DTR$_{\gamma}$ can be therefore obtained by subtracting  the beam monitor contribution DTR$_{m}$, which was measured separately~\cite{andre_2024}. 
DTR$_{m}$ depends strongly on proton energy, while  DTR$_{\gamma}$ only shows a weak beam energy dependence thus  allowing consistent comparison of time resolutions performances across different proton energies. \\
Moreover, TIARA time performances heavily depend on the accelerator time structure. 
%Clinical facilities employ cyclotrons, synchrotrons, and synchro-cyclotrons, each characterized by markedly different  beam structures.
%Output is continuous for cyclotrons, while particles are delivered in spills of a few seconds or a millisecond for synchrotrons and synchrocyclotrons respectively. At a shorter time-scale, they all present a bunched structure with bunch widths going from few to hundreds of ns; at the time-scale of gamma detector ($\mathcal{O}(ns)$), the beam will therefore appear either bunched or continuous, posing different constraints for the detectors. \\
The TIARA module has been tested with proton beams of three clinical accelerators, which are characterized by markedly different beam structures (see table~\ref{tab:beam_structures}): the MEDICYC cyclotron and the the IBA synchro-cyclotron S2C2 (ProteusONE) at the Antoine Lacassagne Centre (CAL) in Nice and the synchrotron at the National Centre for Oncological Hadrontherapy (CNAO) synchrotron in Pavia. MEDICYC provides 63~MeV protons delivered every 40~ns in  bunches of 4~ns width \cite{Hofverberg_2022}. ProteusONE produces protons with energies from 100 to 225~MeV in bunches of 16~ns period and 50\% duty cycle \cite{de_walle_2016}. Finally, the CNAO \cite{rossi_2011_cnao} synchrotron supplies protons (and carbon ions) with energies up to 250~MeV with a pulsed structure that depends on particle energy. During our experiments, we measured a bunch duration of approximately 100~ns with a period of 600~ns; as a matter of fact, at the gamma-ray detection module time-scale ($\mathcal{O}(ns)$), the beam appears continuous.\\

\begin{table}[!t]
    \caption{Main characteristics of proton beams for the three accelerators involved in this study. The CNAO synchrotron allows for variable time periods and bunch duration but only the parameters employed in this study are reported.  CNAO synchrotron and ProteusOne also display a pulsed structure at the s and ms level, respectively, while cyclotrons deliver beams continuously.}
    \centering
    \renewcommand{\arraystretch}{1.3}
    \begin{tabular}{lccc}
        \toprule
	   
	  & MEDICYC & ProteusONE & CNAO  \\
	  & cyclotron & synchro-cyclotron & synchrotron  \\
	   
	  \cmidrule(lr){2-4} 
    	   Energy (MeV) & 63 & 100 -- 225 & up to 250 \\
	  \cmidrule(lr){1-4} 

          Period (ns) & 40 & 16 & $\sim$600 \\
          Bunch width (ns) & 4 & 8 & $\sim$100 \\
 
        \bottomrule
    \end{tabular}
    \label{tab:beam_structures}
\end{table}

In order to tag in time the exact proton generating the PG, pile-up in the beam monitor should be avoided. The beam intensity was therefore reduced to Single Proton Regime (SPR) \cite{Dauvergne_2020}): for beams that are temporally pulsed at the gamma detector scale (MEDICYC and ProteusOne), this corresponds to approximately 1~proton/bunch; at CNAO, the intensity was reduced to $6 \times 10 ^{6}$~p/s.\\
%
%Two types of targets are used: a thin target (a 10~cm thick PMMA-based or a 1~mm thick copper-based) to generate a point-like gamma source and determine the CTR of the detection system; and a thick PMMA target, to stop the beam and measure the full TOF distribution from beam entrance to the Bragg peak. The TIARA modules are placed off-axis, at varying positions and angles with respect to the target (c.f. Fig.~\ref{schema_type}): upstream and/or downstream, in correspondence to the Bragg peak position. 
Finally, background measurements were performed by removing the target in order to quantify scattered protons or PGs generated in the beam monitor among other possible contributions. When necessary, background data is renormalised by the acquisition time and then subtracted from data obtained with the target. \\
%At the low count rates ($\sim$ Hz) typical of SPR, we may safely assume that the acquisition system has no dead time.
%
%
Results obtained at different accelerator facilities are presented in the next subsections, together with the specific characteristics of each experimental setup. 
The three measurements were conducted at different times, thus explaining minor discrepancies in the detector versions employed and their positions with respect to the target.
\subsection{MEDICYC cyclotron -- 63~MeV protons}\label{section_ctr_medicyc}
\begin{figure*}[!h]
    \centering
    \includegraphics[width = \linewidth]{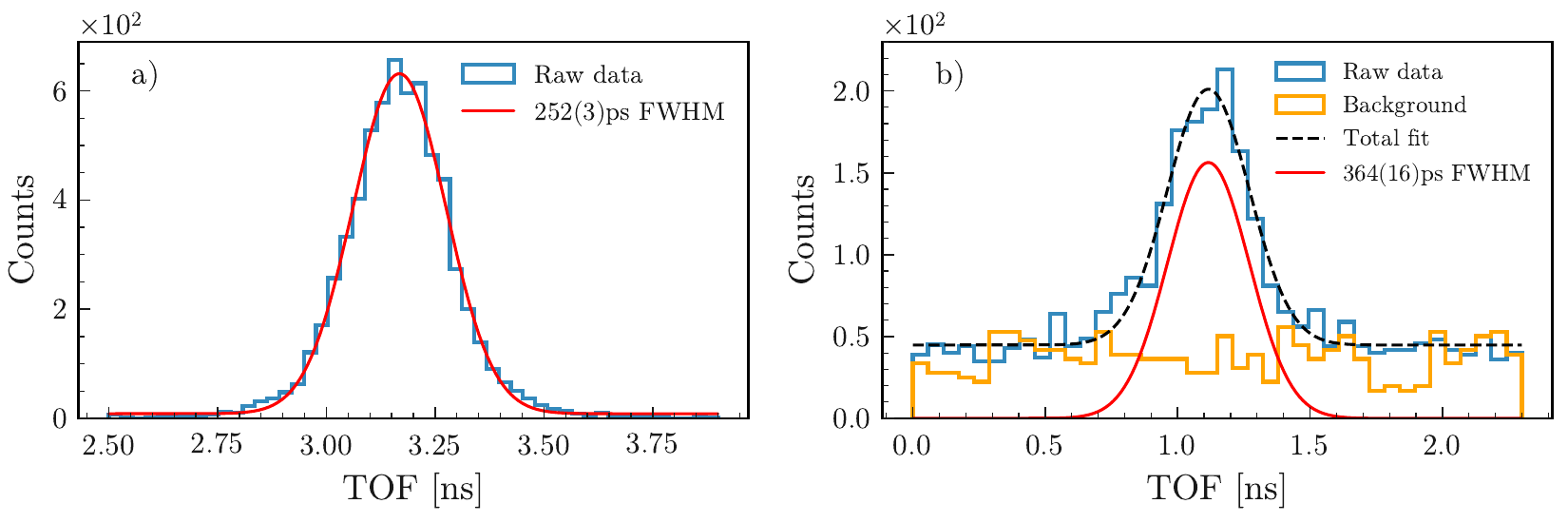}
    \caption{TOF profiles obtained by irradiating a 1~mm copper target at MEDICYC (a) and a 10~mm PMMA slab at CNAO (b). For the CNAO result, raw data (in blue) are superimposed to the flat background (in orange) measured by removing the target. In each plot, the Gaussian fit (red) allows determining the system CTR.}
    \label{CTR_MEDICYC_CNAO}
\end{figure*}

The measurement at MEDICYC was performed by irradiating a 1~mm thick copper target with 63~MeV protons. The final version of the TIARA module (described in Section \ref{sec:det_des}) was placed upstream of the target at approximately 17~cm from the beam axis. The TOF profile between the two detectors is shown in Fig. \ref{CTR_MEDICYC_CNAO}a); these are raw data and the background can be considered negligible. 
The CTR obtained, corresponding to the width of the Gaussian distribution, is $252\pm3$~ps FWHM. Given the beam monitor DTR of $120\pm1$~ps FWHM at 63~MeV, the intrinsic DTR of the TIARA module is estimated at $222\pm3$~ps FWHM. 
\subsection{CNAO synchrotron -- 100~MeV protons}
The final version of the TIARA modules was characterised at CNAO with a 1~cm thick PMMA target irradiated by 100~MeV protons.
 The TIARA module was placed upstream the target at 18~cm from the beam axis. The TOF profile obtained is presented on Fig. \ref{CTR_MEDICYC_CNAO} b). The raw data (in blue) are superimposed to the background (in orange). The latter displays a flat distribution and is essentially determined by false coincidences.
The distribution width is of $364\pm16$~ps FWHM but it does not correspond to the system CTR. 
The 1~cm PMMA target, in fact, was useful to increase the measurement count rate but it can not be considered a point-like source. With a Monte Carlo simulation of this set-up, the proton travel time in the target was estimated to $104 \pm 3$~ps FWHM (including beam energy dispersion) and deconvolved from the experimental data, resulting in a CTR of $349\pm16$~ps FWHM. With the beam monitor DTR separately measured at $204\pm3$~ps FWHM, the gamma-ray detection module DTR is $283\pm16$~ps FWHM. 

\subsection{ProteusONE synchro-cyclotron -- 148~MeV protons}\label{CTR_P1}
\begin{figure*}[!h]
    \centering
    \includegraphics[width = \linewidth]{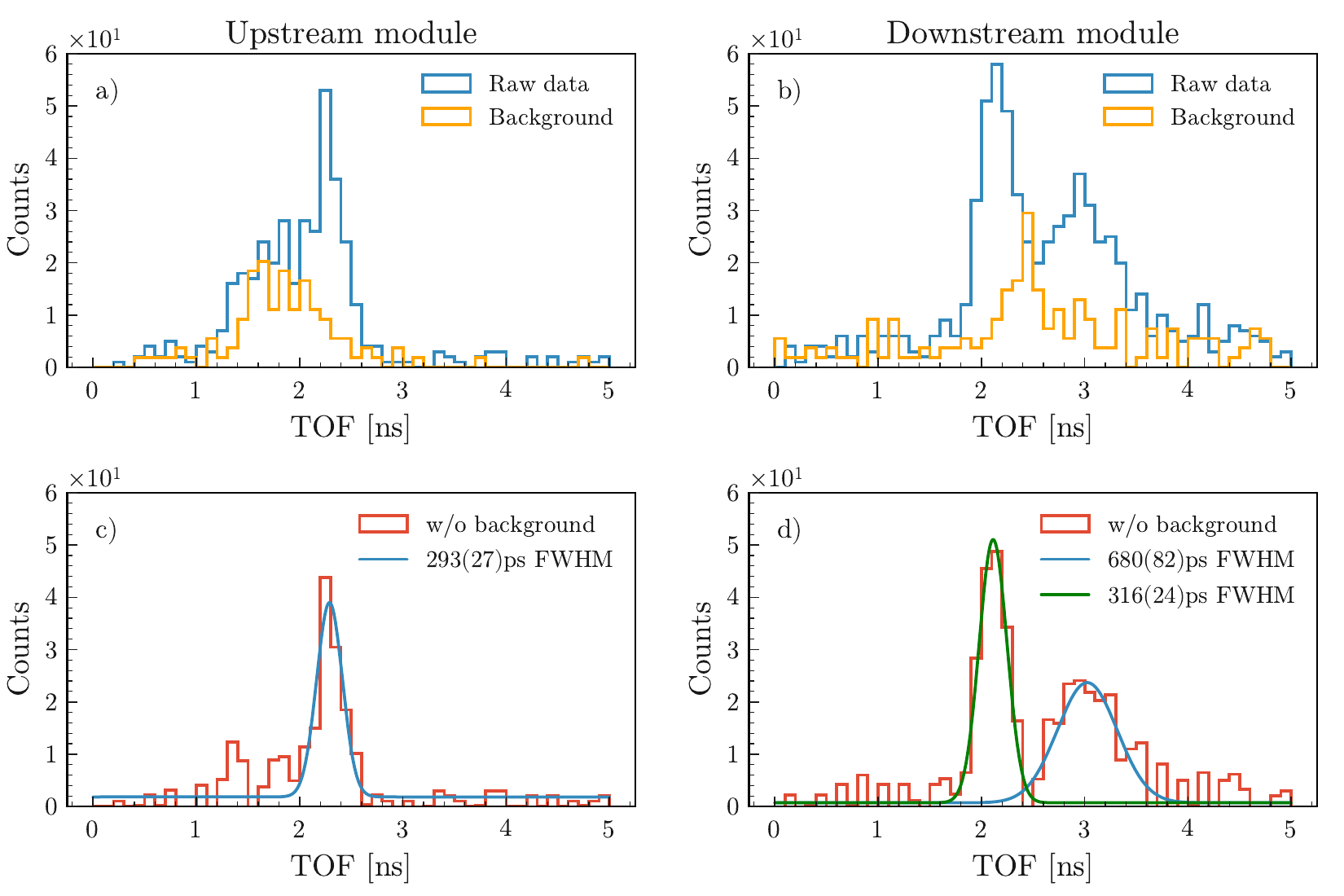}
    \caption{TOF profiles obtained by irradiating a thin target at the ProteusOne accelerator for TIARA modules placed upstream (left) and downstream (right) the target. On  top, raw data are superimposed with the background  corresponding to protons scattered in the beam monitor. The bottom plots represent the TOF profiles after background subtraction. The contribution of PGs emitted in the target is fitted in blue and, for the downstream module, the second contribution around 3~ns corresponds to protons scattered in the target.}
    \label{P1}
\end{figure*}

The ProteusOne campaign was carried out with an earlier version of the detection system: the PbF$_2$ crystal in the TIARA module was smaller ((1.5~cm)$^3$), and the beam monitor was thicker (3~mm instead of 1~mm). Two gamma-ray detection modules were employed, placed upstream and downstream of a 1~mm copper target, irradiated with 148~MeV protons.
Fig. \ref{P1} represents the TOF measured with (in blue) and without (in orange) the thin target for the upstream (a) and downstream modules (b).
Both datasets display an important background contribution; according to Monte Carlo simulations (described in section~\ref{sec:simu_background}) this comes from protons scattered in the beam monitor. 
After background subtraction (Fig. \ref{P1}c and d), a narrow peak generated by PGs emitted in the target (around 2.1~ns) is clearly distinguishable and can be fitted with a Gaussian function. In the downstream module (Fig. \ref{P1}d), a second peak appears at $\sim$3~ns, corresponding to protons scattered from or produced within the target. The detection of these protons does not interfere with the determination of the TIARA module DTR, which is performed for calibration purposes only. In the real applications, however, the target (patient) absorbs both primary and secondary protons. 
The CTR obtained are respectively $293\pm27$~ps FWHM and $316\pm24$~ps FWHM for the upstream and the downstream gamma detectors. With a monitor DTR of $144\pm2$~ps FWHM, the upstream and downstream TIARA modules DTR are $255\pm27$ and $281\pm24$~ps FWHM, respectively. 
\section{Detection efficiency}\label{sec:simu}
The final TIARA detection system will be composed of 30~PG detection modules homogeneously arranged around the target. For particles converted in the crystals, its global detection efficiency $\epsilon$ can be described as:
\begin{equation}
\label{eq_epsilon}
\epsilon = \epsilon_{geo} \times \epsilon_{int} \times \epsilon_{opt}
\end{equation}
with $\epsilon_{geo}$, the solid angle covered by the 30~PG modules;
$\epsilon_{int}$, the probability for a given particle reaching the detector to interact in the PbF$_{2}$ crystal;
and $\epsilon_{opt}$, the probability that the number of Cherenkov photons detected by the SiPM is above the detection threshold (around 8~photoelectrons for measurements listed in this paper).
The last two terms depend on particle's nature and energy, while the first depends on the detector's position and size.\\
In the case of charged particles directly impinging on the SiPM \cite{marrocchesi_2014,carnesecchi_2022}, equation~\ref{eq_epsilon} takes the form:
 \begin{equation}
\label{eq_epsilon_cc}
\epsilon' = \epsilon_{geo}' \times \epsilon_{SiPM} 
\end{equation}
where $\epsilon_{SiPM}$ is the probability that a charged particle is directly detected by the SiPM, and $\epsilon_{geo}'$ depends on the SiPM active area rather than crystal volume.\\
 TIARA detection modules do not allow precise energy measurement since their limited volume prevents the full absorption of PG rays, thus particle type discrimination is not possible. In addition, the clinical proton sources available are not fully characterized at the low intensities relevant to this study, precluding the direct evaluation of particle count rate. Monte Carlo methods are therefore employed to complement the experimental data and provide estimates of both detection efficiency and  background sensitivity.\\
The Geant4 toolkit~\cite{AGOSTINELLI_2003, allison_2016} version 10.04.02 with the physics list QGSP\_BIC\_HP \cite{WRONSKA_2021} is used to assess these properties in two steps. First, the $\epsilon_{opt}$ term of a single module is determined for different particle types and energies by reproducing the optical properties of the Cherenkov radiator from photon generation to their detection in the SiPM (cf. section \ref{simu_opt}). Then, the full TIARA detector is reproduced to obtain the product $\epsilon_{geo} \times \epsilon_{int}$; in this phase, all particles reaching the PbF$_2$ crystals are stored and the output is convolved with $\epsilon_{opt}$ during post analysis. 
%
%%%%%%%
\subsection{Simulation of optical efficiency, $\epsilon_{opt}$}\label{simu_opt}
%%%%%%
%
A single TIARA module consisting of a 2~$\times$ 1.5~$\times$ 1.5~cm$^{3}$ PbF$_{2}$ crystal readout by 4~SiPMs on the back
is simulated with Geant4 in order to establish its detection efficiency for gamma-rays, electrons, protons and neutrons. \\
The classes G4Cherenkov, G4OpAbsorption, and G4OpBoundaryProcess are added to handle the generation, transport, and interactions of Cherenkov photons within the crystal. Optical properties of the monolithic PbF$_{2}$ crystals are defined in the energy range from 1 to 5~eV, corresponding to photons with wavelengths from 248 to 1240~nm. This wavelength range fully covers the detection window of the SiPMs (Hamamatsu 6075PE), whose detection efficiency is extracted from the detector datasheet.
Crystal density is set to 7.77~g/cm$^{3}$, while the PbF$_{2}$ refractive index is taken from the \textit{refractiveindex.info} database \cite{polyanskiy_2024}, and extrapolated for photons energies up to 5~eV. 
Crystal transmittance is based on measurements reported in \cite{Cemmi_2022} for a sample wrapped in a reflective material (such as Mylar). Finally, the reflectivity of the five painted surfaces was taken from the EJ-510 paint datasheet \cite{EJ-510} and extrapolated within the energy range of interest.
The active surfaces of the SiPMs are modelled as four silicon blocks, each measuring 6~$\times$ 6~mm$^{3}$, and spaced by 1.85~mm and 0.85~mm along the y and z axes, respectively, to reproduce the dead space in the matrices. 
The refractive index of the expoxy resin protecting the SiPM sensitive surface is set to 1.55 (Hamamatsu datasheet), and the absorption length to 1~$\times$ 10$^{6}~\mu$m in order to stop the propagation of photons entering the SiPMs. 
The crystal and the SiPMs are separated by a 0.1~mm layer of glue, for which absorption is neglected and the refractive index is set to 1.5~\cite{dymax_glue}.
In the simulation, two types of information are recorded: the interaction of the incident particle within the crystal (particle type, incident energy, deposited energy, event number) and the photons reaching the SiPMs (photon energy and corresponding event number).\\
The SiPM PDE is applied during post-analysis by taking into consideration the photon energy. 
Each photon seen by the SiPM may also generate a \textit{crosstalk} photon. The \textit{crosstalk} probability was set to 14\%, with each secondary photon able to generate an additional photon in turn. The value given in the SiPM datasheet is 7\% for a bare SiPM, but coupling it with a crystal significantly increases this probability, due to \textit{crosstalk} photons reflecting within the crystal and triggering another SiPM cell. This increase depends on the size and optical properties of the crystal coupled to the SiPM, as well as to the applied OverVoltage OV, and is of a factor 2 to 3 at 2–3~V OV \cite{Gola_2014}.
Since this increase in \textit{crosstalk} could not be measured directly in the laboratory, the value was slightly adjusted so that the simulation reproduces the measured detector response (see Section "Simulation calibration", below). 
Finally, the probability of detecting a \textit{dark count} photon within the signal integration window is about 6.4\% per event (from Hamamatsu datasheet) and was considered negligible compared to the \textit{crosstalk}, which occurs at 14\% per detected photon, rather than per event.
\subsubsection*{Simulation calibration}\label{simu_calib}
\begin{figure}[!h]
    \centering
    \includegraphics[width = 8 cm]{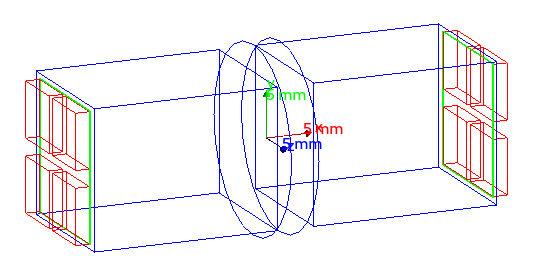}
    \caption{Visualisation of the optical simulation performed with Geant4. Two TIARA modules, each composed of a $2 \times 1.5 \times 1.5$~cm$^{3}$ PbF$_2$ crystal and four $6 \times 6 $~mm$^{2}$ SiPMs interfaced by 0.1~mm of optical grease, are placed on either side of a cobalt source.  The same set-up was realised experimentally to calibrate  the optical parameters in the simulation.}
    \label{simu_cobalt}
\end{figure}
\begin{figure}[!h]
    \centering
    \includegraphics[width = \linewidth]{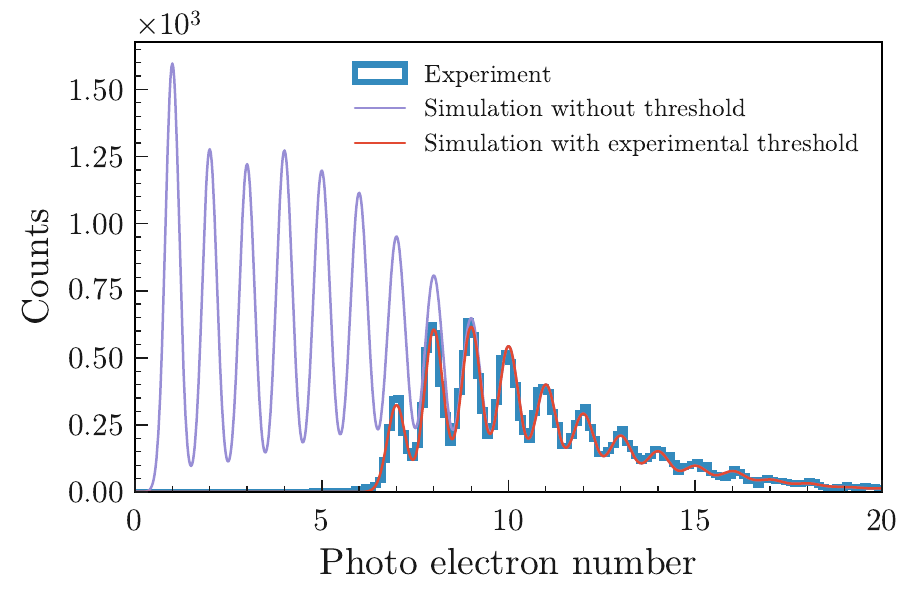}
    \caption{Number of  photo-electrons detected per $^{60}$Co disintegration. The three histograms represent the experimental data (in blue) alongside simulation results in their raw form (in purple) and after the experimental threshold is applied (in red).}
    \label{cobalt_results}
\end{figure}

In order to validate the crystal optical simulation, the same set-up was realised both in Geant4 and experimentally: the simulated parameters reported above have been fine-tuned by adjusting the simulation to experimental data.\\ 
Two identical TIARA modules consisting of a 2~$\times$ 1.5~$\times$ 1.5~cm$^{3}$ PbF$_{2}$ crystal readout by 4~SiPM on the back are irradiated by a $^{60}$Co source emitting two gamma-rays of 1.17 and 1.33~MeV almost simultaneously (cf. Fig.~\ref{simu_cobalt}).
The simulated source is a cylinder with a radius of 1.5~mm and a height of 0.5~mm located in the center of a plastic cylinder with a radius of 10~mm and a height of 1.5~mm to reproduce its spatial spread and the attenuation of its packaging (mainly for electrons). The simulated geometry is represented on Fig.~\ref{simu_cobalt}.\\
At this low gamma-ray energy the number of detected photons is small but may still be counted by integrating the signal in a $\sim$8~ns window, for comparison with the simulation output. In Fig.~\ref{cobalt_results}, experimental data (in blue) are compared to simulated data that are convolved with Gaussian functions reproducing the spread introduced by the electronic chain (in purple).
Count statistics below 10~photoelectrons are biased by the detection threshold applied to experimental data: this consists in a voltage value, while photons are counted from the signal integral. From comparison of simulation and experiments, we could establish that, for the applied threshold, only 34, 75 and 95\% of events with 7, 8 and 9~photons are acquired, respectively. Such a high threshold may seem conservative, but it is necessary to effectively suppress the SiPM dark count ($\sim$1~kHz at 6 p.e.) when acquiring data in single proton regime (this study) for which the PG count rate is between 2 and 4~kHz, depending on proton energy. At nominal beam intensity, the PG count rate is of the order of the MHz and the acquisition threshold can be lowered.

\subsubsection*{Results}
\begin{figure*}[!h]
    \centering
    \includegraphics[width = \linewidth]{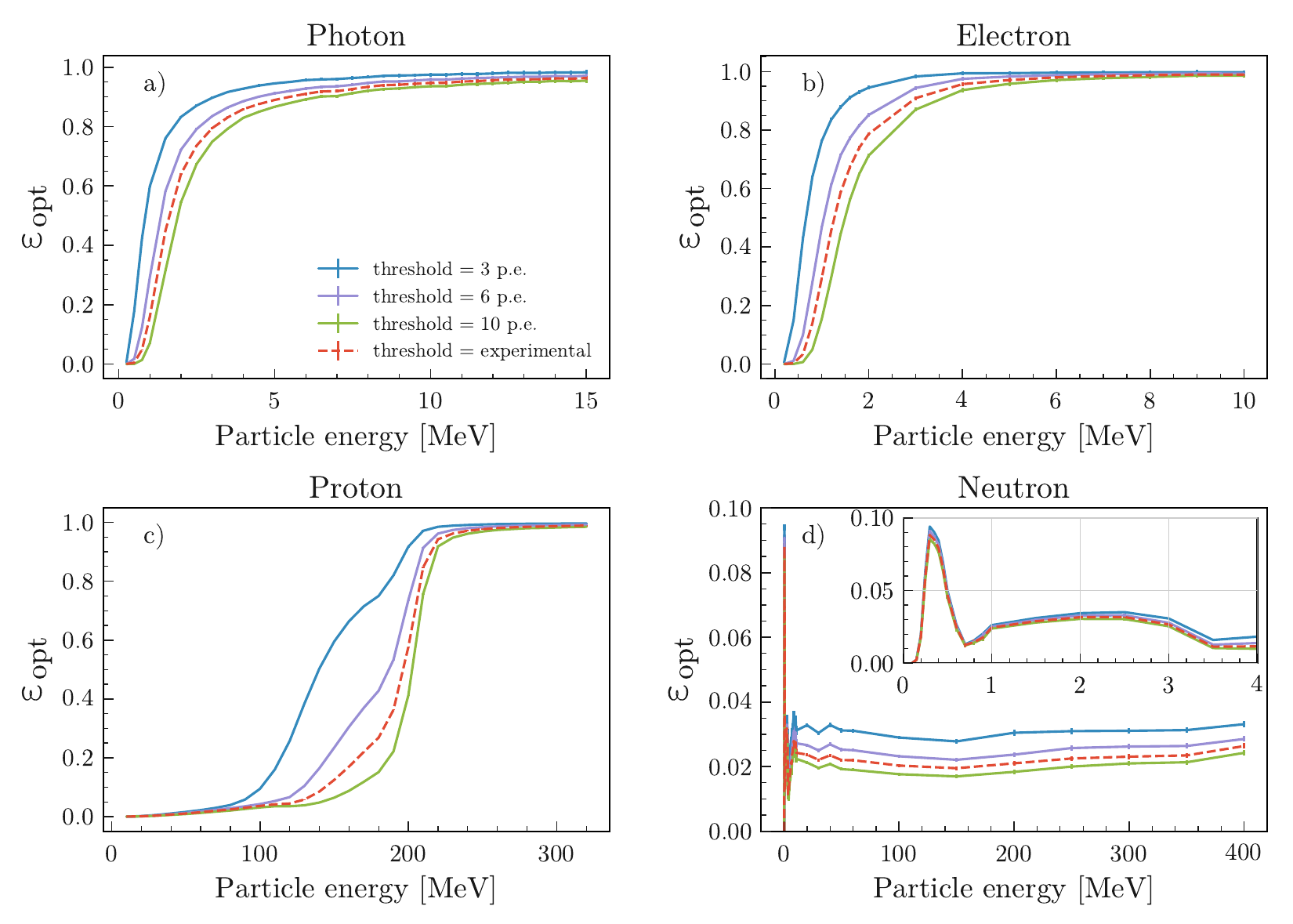}
    \caption{Optical detection efficiency of the TIARA module as a function of the kinetic energy of the incident particle, for (a) photons, (b) electrons, (c) protons and (d) neutrons. Results for fixed detection thresholds expressed in number of photo-electrons (p.e.) are reported, as well as those obtained with the experimental detection threshold (dashed line).}
    \label{det_eff}
\end{figure*}

Once calibrated, the optical simulation was used to quantify the optical detection efficiency ($\epsilon_{opt}$) of the TIARA module for photons, electrons, protons, and neutrons over a range of relevant energies (cf. Fig.~\ref{det_eff}). A single module was simulated facing a beam with varying particle type and incident energy. Detection efficiencies were evaluated for four thresholds: three fixed values (3, 6, and 10~p.e.) plus the experimental threshold determined with the $^{60}$Co calibration measurement described above.
For gamma-rays, simulations are performed from 0.1 to 15~MeV. The detection efficiency decreases sharply below $\sim$2~MeV, which is advantageous as it suppresses annihilation photons (511~keV) and delayed gamma rays (e.g., 0.718~MeV from $^{10}$B with mean emission time of 27.8~s \cite{kozlovsky_2002}) that are not time-correlated. Above 2~MeV, the efficiency rises steeply, reaching ~90\% for high-energy PGs.\\
For electrons, efficiencies are studied between 0.1 and 10~MeV. The efficiency profile resembles that of photons, with nearly 100\% efficiency above 2~MeV and a drop at lower energies. This cut-off is beneficial as it rejects low-energy electrons produced in air, that are not temporally correlated. Unlike photons, electrons always interact in the crystal ($\epsilon_{int}$ = 100\%), while photon interactions follow an exponential attenuation law.\\
For protons, $\epsilon_{opt}$ was evaluated from 10~to 320~MeV. Although most clinical proton beams do not exceed 225~MeV, higher energies are relevant for secondary protons in carbon therapy, where TIARA could also be applied. The efficiency is 100\% above 225~MeV, indicating strong sensitivity to secondary proton background. The Cherenkov threshold for PbF$_2$ ($n$~=~1.8) corresponds to 190~MeV, explaining the sharp efficiency increase above this value. Below 190~MeV, Cherenkov photons arise only from secondary electrons. With the experimental threshold, detection of scattered protons between 100~and 200~MeV is limited, and efficiency below 100~MeV is nearly zero, which is consistent with earlier observations of reduced background at 63~MeV (MEDICYC) irradiation compared to 148~MeV (ProteusOne) irradiation.\\
For neutrons, efficiencies were calculated from 0.15 to 400~MeV, covering the range of secondary neutrons produced in proton and carbon therapies. Neutrons are a major source of background for PGTI because their delayed emission leads to some overlap with the fall-off of the PG TOF distribution, degrading the Bragg peak localization measurement. However, the use of a pure Cherenkov radiator makes TIARA nearly insensitive to direct neutrons detection. Applying the experimental detection threshold, neutron efficiency is limited to $\sim$2\% between 1 and 400~MeV, slightly increasing ($\sim$8\%) around 0.3~MeV, possibly due to changes in cross-sections or Geant4 physics models. These low-energy neutrons are not temporally correlated and can be rejected by TOF selection. Nevertheless, gamma rays that are indistinguishable from the PG signal may still be produced in nuclear reactions of fast neutrons with carbon or oxygen nuclei in the patient.
%Gamma rays are also be produced in nuclear reactions of fast neutrons with the target, e.g., in (n,n’), (n,p), or (n,alpha) reactions with carbon or oxygen nuclei. Such gamma rays, and not (only) the direct detections of fast neutrons, are responsible for the “non-linear” background in PGT measurements.
%
\subsection{Direct detection of charged particles, $\epsilon_{SiPM}$}\label{sec:simu_background}
The TIARA module signal is not only due to particles that are converted in the crystal, but also, in minor part, to charged particles (electrons, positrons and protons) directly interacting in the SiPM \cite{marrocchesi_2014,carnesecchi_2022}. 
%Among these particles, electrons, positrons, and protons should be considered. 
Positrons and a fraction of electrons generated within the crystal by PG interactions are excluded, as PG detection is already accounted for in the optical efficiency. The remaining electrons are produced in air along the proton path, with energies below 0.5~MeV in 99\% of cases (as from Monte Carlo simulation); the 1~mm thick aluminium packaging used to shield the SiPM and its electronic is sufficient to absorb these particles \cite{nist}.
For protons above a few keV, the detection efficiency of SiPMs is equivalent to their fill factor \cite{Ogasawara_2017}: this corresponds to a $\epsilon_{SiPM}=0.82$ for the Hamamatsu S13360-6075PE SiPMs mounted in the PG module. 
\subsection{Background sensitivity}\label{sec:back}
\begin{figure*}[!h]
    \centering
    \includegraphics[width = \linewidth]{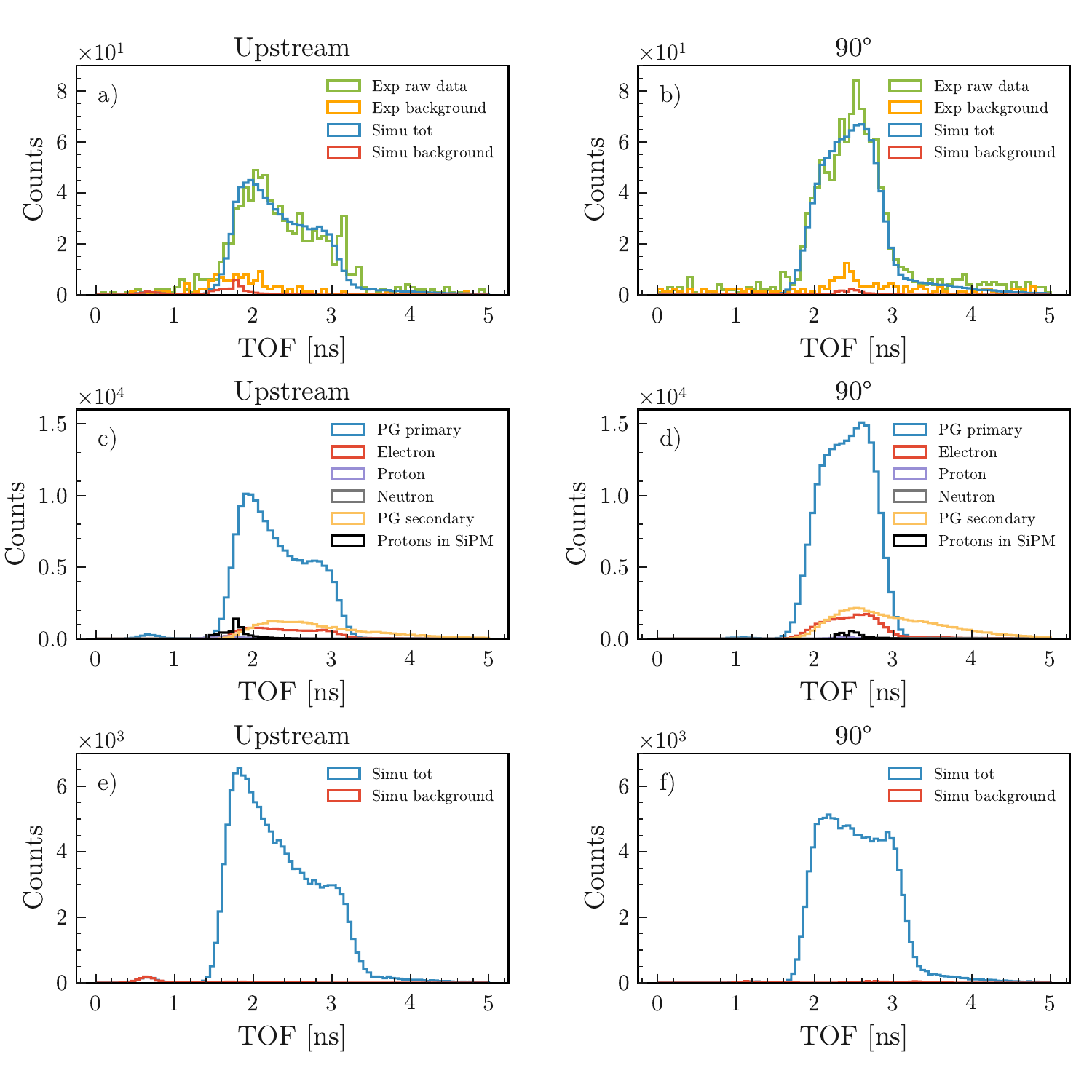}
    \caption{Top: Irradiation  of a thick PMMA target with 148~MeV protons. TOF's are measured between the 3~mm thick plastic monitor and the TIARA modules placed at 13~cm from the beam axis. Simulated and experimental data are compared for modules placed (a) upstream and (b) at 90° with respect to the Bragg peak position along the beam axis. For simulated data, TOFs  are convolved with the experimentally determined CTR.
    Centre: In (c) and (d),  the separate contributions of different particle types are reported for both modules. The background is mainly due to scattered protons detected directly by the SiPMs. 
    Bottom: A simulation closer to the final TIARA design  shows that the background contribution is significantly reduced for both detectors (e and f) when employing a thinner beam monitor (1~mm)  and moving away the gamma-ray detection modules from the beam axis (25~cm).}
    \label{simu_background}
\end{figure*}

During the ProteusOne campaign described in section \ref{CTR_P1}, the 1~mm copper target was replaced with a PMMA block thick enough to fully stop 148~MeV proton beam. The experimental set-up is the one described in Fig.~\ref{schema_type}, with the two gamma detectors positioned at 13~cm from the beam axis. With the thick target, the module previously considered downstream is now positioned at 90° with respect to the Bragg peak position along the beam axis. Preliminary prototypes of both the beam monitor (3~mm thick) and TIARA detector ((1.5~cm)$^3$ volume) were employed. 
The TOF distributions obtained are presented in Fig.~\ref{simu_background} c) and d) for modules placed upstream the target and at 90° with respect to the Bragg peak position, respectively. Background data obtained without the target are also displayed. \\
The same experiment was reproduced in a Geant4 with the aim of identifying and quantifying the background signal of TIARA. 
The simulation allows counting the fraction of particles converted in the crystals ($\epsilon_{geo} \times \epsilon_{int}$ in Eq.~\ref{eq_epsilon}) and the amount of charged particles reaching the SiPM ($\epsilon_{geo}'$ in Eq.~\ref{eq_epsilon_cc}), for each particle type and energy. The two functions are then convolved by $\epsilon_{opt}$ and $\epsilon_{SiPM}$, respectively, in order to obtain the actual number of detected particles. The separate contributions per particle type is reported in Fig.~\ref{simu_background} c) and d) while in figures a) and b) their sum is superposed to experimental data.\\
In general, the TOF distributions were found to vary strongly with the position of the PG module. This effect arises from a combination of solid angle differences and the dependence of PG TOF on detector position. Upstream detectors collect more PGs emitted at the entrance of the beam in the target (around 2~ns), while downstream detectors detect more PGs emitted in proximity of the Bragg peak (2.5-3~ns). In addition, TOF distributions broaden for upstream detectors as deeper PG emission increases both the proton and the PG travel times, whereas for downstream detectors, this effect partly cancels, yielding to narrower spectra. Overall, upstream placement is preferable in PGT, as it enhances TOF separation between emission depths. However, in a clinical setting with a rotating gantry, the upstream/downstream configuration is inconsistent.\\
Regarding background, simulated noise is underestimated compared to measurements (by factors of $\sim$4 and $\sim$7 for upstream and downstream modules, respectively). This discrepancy may result from uncertainties in beam dispersion, simplified modelling of the monitor, or variations in secondary particle production cross-sections in Geant4 physics lists. Differences in the response among module samples and versions, as well as threshold variations linked to SiPM gain and temperature may also contribute. Nonetheless,  unlike thin-target measurements, thick (patient-like) targets result in a high SNR of $\sim$10. This improvement arises from the larger number of PGs emitted and the reabsorption of secondary protons within the target.\\
Still, simulations allow identification of the signal and background components. Simulated TOF spectra per particle type (Fig.~\ref{simu_background}~c) and~d)) show that the majority of the signal originates from primary PGs. Secondary PGs contribute only with a small tail that degrades information near the end of the proton range; this component includes gamma rays generated from nuclear reactions of fast neutrons with oxygen and carbon nuclei.
A minor contribution from electrons is observed, but their TOF distribution matches that of primary PGs, thus increasing the statistics of the relevant signal. Neutron background from direct detection is negligible. Finally, protons detected directly by the SiPMs account for most of the noise observed in measurements without a target.
Unfortunately, it is not possible to discriminate these protons from PG signals experimentally, but their contribution can be limited by reducing the thickness of the beam monitor (from 3~mm to 1~mm, as with our final prototype) and by moving the detectors away from the beam axis.
%
%%%%
\subsection{Sensitivity}
%
%
%Through simulation it is also possible to estimate  the overall sensitivity of the TIARA system, defined as the expected number of detected events per incident proton. 
%The efficiency accounts for the different contributions given in equations~\ref{eq_epsilon} and  \ref{eq_epsilon_cc}, as well as the PG production rate as given by the QGSP\_BIC\_HP physics list in Geant4. Results are listed in Table~\ref{tab:stat} for the three energies considered in this paper. 

%%%%%%%%
To assess the performance of the future TIARA detection system composed of 30~modules, a simulation was conducted using a geometry closer to the final design. The setup included a 1~mm thick monitor and 30~gamma-ray detection modules placed on a 25~cm radius sphere centred on a 10~cm water target.
Two of the 30~modules were positioned at 45° and 90° relative to the beam direction for comparison with the experimental data described in the previous section.\\
The resulting TOF distributions, shown in Fig.~\ref{simu_background} e) and f), indicate that reducing the monitor thickness and increasing the gamma-ray detection module distance improves the SNR by a factor of 1.7 (for the module at 90°) compared to the previous geometry: the proton background in the SiPMs becomes negligible for both the detectors, while PGs emitted in the monitor (at $\sim$0.6~ns for the upstream module) can be temporally discriminated. \\
%Therefore, the expected SNR for the final TIARA system is 17.\\
%
With this set-up, it was also possible to estimate the term $\epsilon_{geo} \times \epsilon_{int}$ that, convolved to the  optical efficiency $\epsilon_{opt}$ determined previously, gives an overall detection efficiency for gamma-rays of 0.45\%.\\ 
Finally, the sensitivity of TIARA, defined as the expected number of detected events per incident proton, was determined. These values, which are listed in Table~\ref{tab:stat} for the three energies considered in this paper, account
for the separate contributions of the different particles detected by the modules, as well as the PG production rate as given by the QGSP\_BIC\_HP physics list of Geant4. %Results are listed in Table~\ref{tab:stat} for the three energies considered in this paper. 

\begin{table}[!t]
    \caption{Expected number of PGs emitted and events detected with the final TIARA system for a single spot of $1\times10^{7}$ incident protons and the three proton energies considered in this study. Values are obtained from Geant4 simulations implementing the QGSP\_BIC\_HP physics list.}
    \centering
    \renewcommand{\arraystretch}{1.3}
    \begin{tabular}{lccc}
        \toprule
        \makecell{Proton energy} & 63~MeV & 100~MeV & 148~MeV \\
        \cmidrule(lr){2-2} \cmidrule(lr){3-3} \cmidrule(lr){4-4}
        \makecell{Nb.  incident\\ proton} & $1\times10^{7}$ & $1\times10^{7}$ & $1\times10^{7}$ \\
        \cmidrule(lr){2-2} \cmidrule(lr){3-3} \cmidrule(lr){4-4} 
        \makecell{Nb.  PG\\emitted} & $3.5\times10^{5}$ & $6.7\times10^{5}$ & $11.5\times10^{5}$ \\
        \cmidrule(lr){2-2} \cmidrule(lr){3-3} \cmidrule(lr){4-4} 
        \makecell{Nb. events \\ detected} & $1.6\times10^{3}$ & $2.9\times10^{3}$ & $4.9\times10^{3}$ \\
        \cmidrule(lr){2-2} \cmidrule(lr){3-3} \cmidrule(lr){4-4} 
        \makecell{Sensitivity} & $0.016$\% & $0.029$\% & $0.049$\% \\
        %\cmidrule(lr){2-2} \cmidrule(lr){3-3} \cmidrule(lr){4-4} 
        %\makecell{Range accuracy} & $2.9\pm0.1$~mm & $4.4\pm0.1$~mm & $9.6\pm0.7$~mm \\
        \bottomrule
    \end{tabular}
    \label{tab:stat}
\end{table}
\section{Range accuracy}\label{sec:range}
\begin{figure}[!h]
    \centering
    \includegraphics[width = 8 cm]{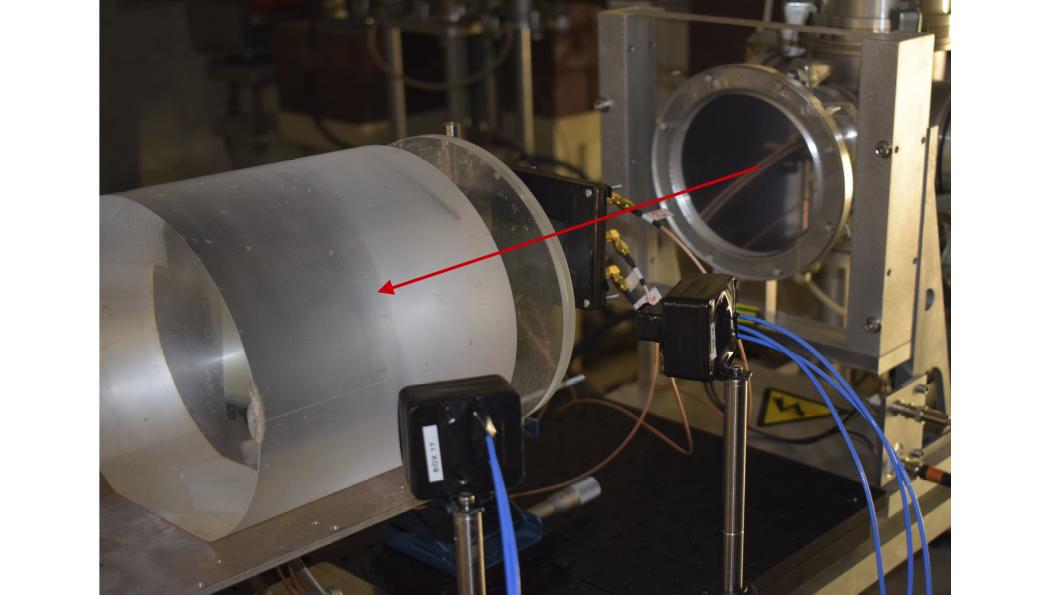}
    \caption{Picture of the spatial sensitivity measurement performed at MEDICYC with 63~MeV protons. The two PMMA targets are initially separated by a variable air gap (from 30 to 40~mm). The red arrow represents the proton beam direction.}
    \label{photo_sensitivity}
\end{figure}

The aim of this experiment was to quantify the spatial accuracy of TIARA for the measurement of a proton range shift with the PGT technique. 
The measurement was performed on the MEDICYC accelerator at 63~MeV and in SPR. The set-up is composed of two PMMA targets, 1~cm thick upstream and 17~cm thick downstream (Fig. \ref{photo_sensitivity}), separated by a 30~mm air gap. The air gap is then progressively increased in steps of 1 or 2~mm by moving the thin target in the upstream direction, in order to produce artificial proton range shifts in the range [0-10]~mm.
 The detectors and their positions are those described in section \ref{section_ctr_medicyc}, with a CTR of 252~ps FWHM. 
\begin{figure*}[!h]
    \centering
    \includegraphics[width = \linewidth]{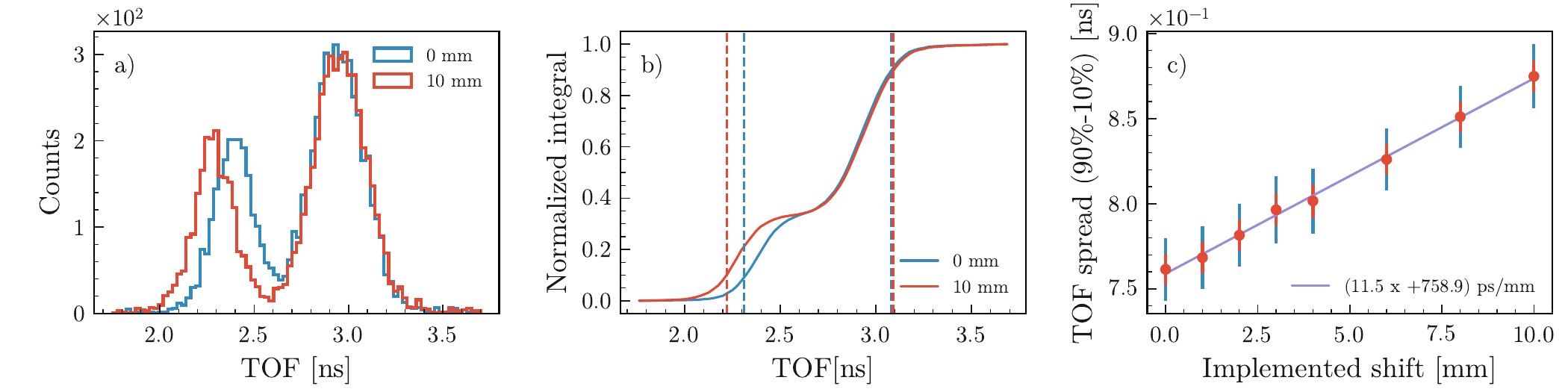}
    \caption{Different steps for the determination of the range accuracy. The experimental TOF distributions (a) and their cumulated integral (b) are reported for the reference position (in blue) and for a 10~mm increase in the air gap between the targets (in red). The proton range is evaluated as the time delay between 10 and 90\%
   of the integral profile (dashed vertical lines). In (c), the mean proton range expressed in time-delay units is plotted against the shift in air gap thickness. Error bars are expressed as 1$\sigma$ (red) and 2$\sigma$ (blue) values. The 2$\sigma$ range accuracy corresponds to the minimum spatial shift for which the  2$\sigma$ bars do not overlap. }
    \label{sensitivity}
\end{figure*}
For each step, the measured TOF profile (Fig.~\ref{sensitivity}, left) is integrated in order to smooth out statistical fluctuations. The cumulative integral (Fig.~\ref{sensitivity}, center) is then employed to evaluate the time delay between 10\% and 90\%, thus providing an indirect measurement of the proton range. 
This measurement is repeated on 1000~sub-samples bootstrapped from the experimental dataset, each consisting of $1.6\times10^{3}$ events, corresponding to the PG statistics expected for a 10$^{7}$ protons spot at 63~MeV (cf. Table~\ref{tab:stat}): 
the resulting distribution allows evaluating the 1$\sigma$ and 2$\sigma$ errors on proton range. \\
Fig.~\ref{sensitivity} a) and b) illustrate the procedure for the initial target position (0~mm shift, in blue) and the maximum shift applied (10~mm in red). 
On the TOF profile (left), the peak around 3~ns, corresponds to PGs emitted in the fixed thick target and it is identical for both geometries. In contrast, the peak around 2.3~ns corresponds to PGs from the thin target and its position varies according to the implemented shift. The right plot shows the measured proton ranges (mean value in time-delay units) for each relative increase of the the air gap; error bars indicate the 1$\sigma$ (red) and 2$\sigma$ (blue) values. From these data, the proton range accuracy can be estimated to $3.3 \pm 0.1$ mm at 2$\sigma$ for a small irradiation spot of $\sim$$10^{7}$ protons: this  corresponds to the minimum spatial shift for which the  2$\sigma$ bars do not overlap.
The accuracy depends on both the system CTR and the measured statistics. Combining the PG statistics obtained with two irradiation spots (2~$\times$ 10$^{7}$ protons and 3.2~$\times$~10$^{3}$ TOFs measured), for example, the sensitivity becomes 2.3~$\pm$ 0.1 at 2$\sigma$. 

\section{Discussion}
This work summarizes the progress made over the past two years on the TIARA modules and their experimental characterization. Assessing the detector response is challenging, since particle energy cannot be measured directly and the low light yield of PbF$_2$ hinders particle discrimination. A combined approach of experimental measurements and Monte Carlo simulations was therefore required to achieve a comprehensive description of their performance.\\
Despite these challenges, the use of pure Cherenkov radiators offers significant advantages over scintillators. The TIARA modules are essentially blind to neutrons, leading to a very high SNR, with the background practically negligible at 25~cm from the beam axis when operated with a bunched accelerator such as ProteusOne (expected SNR = 17). Moreover, the fast Cherenkov signals effectively suppress pile-up, even at the higher intensities typical of clinical treatments (not shown here).\\
The resulting high SNR translates into very good detection efficiency (0.45\% expected for the final TIARA system), with a direct impact on spatial accuracy. In terms of range verification, an accuracy of $3.3 \pm 0.1$ mm at 2$\sigma$ was achieved at 63~MeV, for an irradiation spot corresponding to $\sim10^{7}$ protons. A foreseen application of TIARA is the verification of a few initial spots in SPR at the beginning of treatment (during few tens of seconds\cite{Jacquet_2023}), where the intrinsic time resolution of the system can be fully exploited without being affected by pile-up in the beam monitor.
Once the clinical intensity is reached for the remainder of the treatment, higher counting statistics become available, and a comparable range accuracy is expected \cite{Jacquet_2021}.\\
Range accuracy is closely linked to the CTR of TIARA, which was found to be 252~ps FWHM under favourable conditions but varies across experiments. The contribution from the beam monitor depends strongly on the proton energy, since higher-energy particles display lower specific energy loss, while the TIARA module DTR shows only a weak dependence on beam energy. The best performance was observed at MEDICYC (63~MeV), whereas values up to 283~ps FWHM were recorded at CNAO. These discrepancies are not necessarily structural, as multiple experimental factors varied, including detector version, beam time structure, beam size, and angular dispersion. In particular, the CNAO result can likely be attributed to the use of a thick PMMA target, for which the point-like approximation was no longer valid and required post-analysis correction, as well as to the asynchronous proton arrival times—on the scale of the gamma-ray detector—which complicated proton-gamma coincidence identification.\\

\section{Conclusions}
Overall, the  results presented in this study indicate that the TIARA modules hold promise for improving range verification in proton therapy, although further studies are ongoing to validate their performance under the higher-intensity conditions that are typical of clinical practice.
\section*{Acknowledgments}
All authors declare that they have no known conflicts of interest in terms of competing financial interests or personal relationships that could have an influence or are relevant to the work reported in this paper. 
This work was supported by the European Union (ERC project PGTI, grant number 101040381). Views and opinions expressed are however those of the authors only and do not necessarily reflect those of the European Union or the European Research Council Executive Agency. Neither the European Union nor the granting authority can be held responsible for them.
This work was partially supported by the European Union’s Horizon 2020 research and innovation programme HITRIplus (grant agreement no. 101008548) via the Transnational Access (TNA) framework.

\bibliographystyle{IEEEtran}

\bibliography{bibliography}

\vfill

\end{document}